\documentclass{emulateapj}
\usepackage{topcapt}
\usepackage{booktabs}
\usepackage{natbib}
\usepackage{rotating}
\usepackage{longtable}
\shorttitle{IMF in M83}
\shortauthors{Andrews et al.}

\begin{document}

\title{Big Fish in Small Ponds: Massive Stars in the Low Mass Clusters of M83}

\author{J.E. Andrews\altaffilmark{1}, D. Calzetti\altaffilmark{1}, R. Chandar\altaffilmark{2}, B.G. Elmegreen\altaffilmark{3}, R.C. Kennicutt\altaffilmark{4}, Hwihyun Kim\altaffilmark{5,6}, Mark R. Krumholz\altaffilmark{7}, J.C. Lee\altaffilmark{8,9}, Sean McElwee\altaffilmark{1}, R.W. O'Connell\altaffilmark{10}, B. Whitmore\altaffilmark{8}}

\altaffiltext{1}{Department of Astronomy, University of Massachusetts, Amherst, MA 01003, USA; jandrews@astro.umass.edu, callzetti@astro.umass.edu}
\altaffiltext{2}{Department of Physics and Astronomy, University of Toledo, Toledo, OH 43606, USA}
\altaffiltext{3}{IBM T.J. Watson Research Center, Yorktown Heights, NY 10598 USA}
\altaffiltext{4}{University of Cambridge, Institute of Astronomy, Madingley Road, Cambridge, CB3 0HA, UK}
\altaffiltext{5}{School of Earth and Space Exploration, Arizona State University, Tempe, AZ 85287-1404, USA}
\altaffiltext{6}{Korea Astronomy and Space Science Institute, Daejeon 305-438, Republic of Korea}
\altaffiltext{7}{Department of Astronomy and Astrophysics, University of California, 1156 High Street, Santa Cruz, CA 95064, USA}
\altaffiltext{8}{Space Telescope Science Institute, 3700 San Martin Drive, Baltimore, MD 21218, USA}
\altaffiltext{9}{Visiting Astronomer, Spitzer Science Center, Caltech Pasadena, CA 91125 }
\altaffiltext{10}{Department of Astronomy, University of Virginia, P.O. Box 400325, Charlottesville, VA, 22904-4325, USA}

\begin{abstract}
We have used multi-wavelength \emph{Hubble Space Telescope} WFC3 data of the starbursting spiral galaxy M83 in order to measure variations in the upper end of the stellar initial mass function (uIMF) using the production rate of ionizing photons in unresolved clusters with ages $\leq$ 8 Myr.  As in earlier papers on M51 and NGC 4214, the upper end of the stellar IMF in M83 is consistent with an universal IMF, and stochastic sampling of the stellar populations in the $\lessapprox$ 10$^{3}$ M$_{\sun}$ clusters are responsible for any deviations in this universality. The ensemble cluster population, as well as individual clusters, also imply that the most massive star in a cluster does not depend on the cluster mass.  In fact, we have found that these small clusters seem to have an over-abundance of ionizing photons when compared to an expected universal or truncated IMF.  This also suggests that the presence of massive stars in these clusters does not affect the star formation in a destructive way.
 
\end{abstract}


\section{Introduction}

The blueprint of how stars are formed in galaxies, better known as the stellar initial mass function (IMF), is one of the most essential quantities in astronomy, yet its functional form and universality is still under much debate.    IMF measurements in the nearby Milky Way and Magellanic Clouds have indicated a constant IMF \citep{2011ApJ...739L..46O, 2013ApJ...762..123W}, yet other studies have found evidence pointing towards a non-universal IMF \citep{2011ApJ...735L..13V,2011arXiv1112.3340K,2012Natur.484..485C,2013ApJ...771...29G}.   Variations can be found in the high-mass end (upper IMF; uIMF) and the low-mass end, and both can affect the star formation history (SFH) of a galaxy.  A top-heavy IMF, in which more high-mass stars are formed than predicted from the standard model (for example M82F; \citet{2001MNRAS.326.1027S}), will result in a low mass-to-light ratio and more rapid energy and chemical enrichment of galaxies as stars $>$ 8 M$_{\sun}$ become core collapse supernovae. Variations of the uIMF are also tied to star formation rates (SFRs) and short-timescale SFHs ($<$ 100-500 Myr.) A bottom-heavy IMF, which has an over abundance of low mass stars,  therefore has a high mass-to-light ratio, higher numbers of stars formed (albeit of mostly stars $<$ 1 M$_{\sun}$), and affects estimates of long-term SFHs. The work presented here concentrates on the variations of the uIMF, although discovery of non-uniformity in the high-end would not necessarily discount a variation in the lower limit as well.  

 \begin{figure*}[t!] 
   \centering
   \includegraphics[width=6.3in]{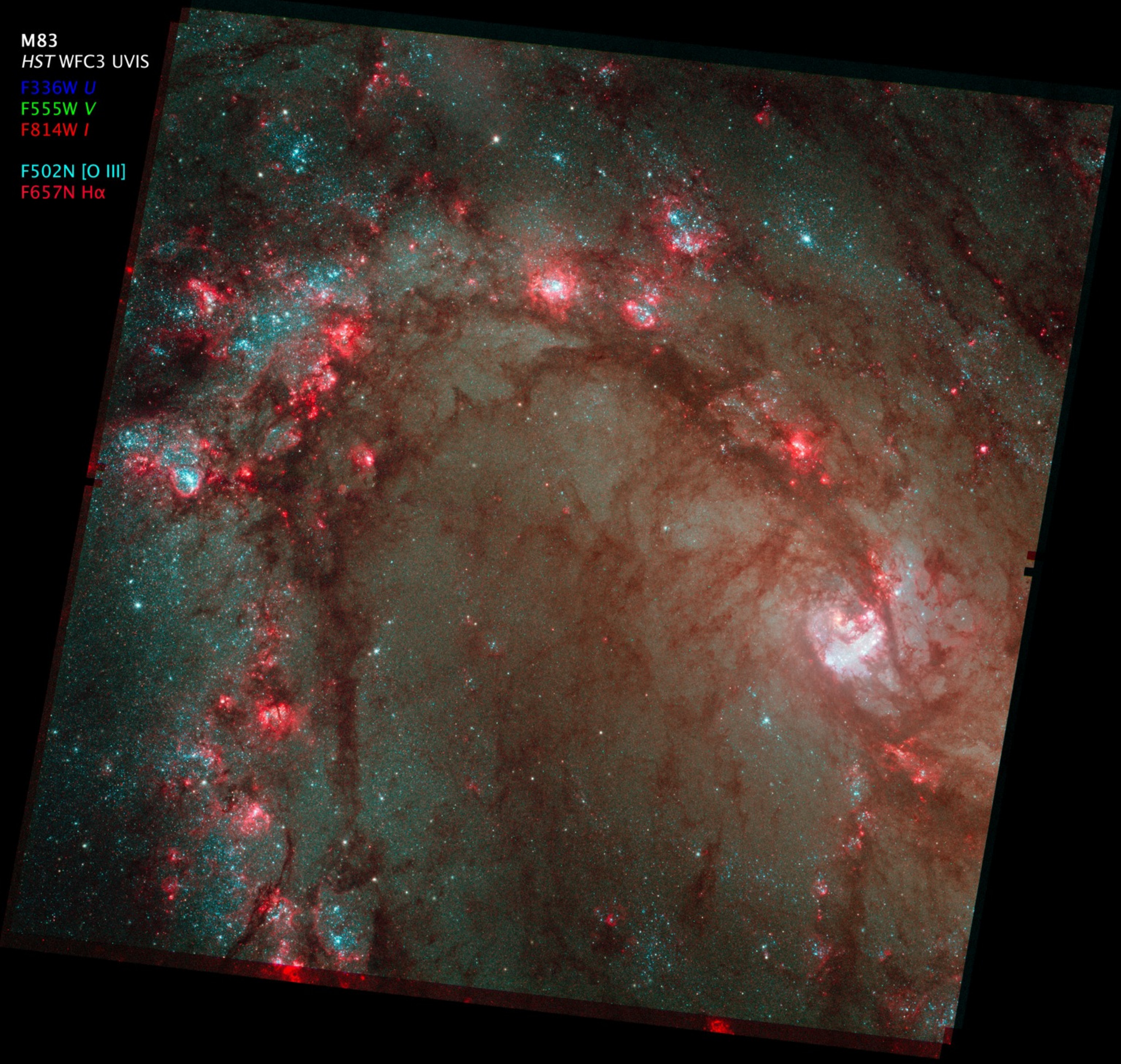} 
   \caption{Color composite WFC3 image of the inner field (F1) of M83 courtesy of Zoltan Levay (STScI-2011-14), R. O'Connell (GO 11360), and the WFC3 SOC. }
   \label{fig:clusterpos}
\end{figure*}

In this paper, we narrow our interest to the lack of ionizing photons per optical or UV luminosity that has been suggested for dwarf starburst galaxies \citep{2008ApJ...675..163H,2009ApJ...706..599L,2009ApJ...695..765M,2009ApJ...706.1527B,2011MNRAS.415.1647G}. Recently, \citet{2011ApJ...741L..26F}, \citet{2012ApJ...744...44W}, and \citet{2012MNRAS.422..794E} have shown that the observed L$_{H\alpha}$/L$_{FUV}$ deviations can be due to bursty star formation or a stochastically populated IMF, removing the necessity for a variant IMF. Observational constraints on this issue are therefore essential to understanding the fundamental evolution of galaxies, especially at high-redshift. For full, comprehensive reviews on this subject we refer you to \citet{2010ARA&A..48..339B}, \citet{2013pss5.book..115K}, \citet{2013arXiv1312.5326O}, and \cite{2014arXiv1402.0867K}.

The IMF can be measured nearby in the Milky Way, Large Magellanic Cloud (LMC), and Small Magellanic Cloud (SMC) by counting the individual stars in clusters young enough ($\leq$ 3-5 Myr) that the most massive stars still remain  \citep[for example]{2000ApJ...533..203S,2008AJ....135..173S,2009ApJ...697L..58A}. 
In addition to the rapid evolution of massive stars, observing clusters at young ages is necessary to get a full census of the stellar population as up to 80$\%$ of stellar clusters experience early mass loss and do not survive longer than 10 Myr \citep[``infant mortality'']{2003ARA&A..41...57L}. Age is not the only problem for individual star counts, as selection biases due to crowding can cause an incomplete sample.  For instance, mass segregation may cause the more massive stars to sink towards the center of the cluster, where it will be harder to distinguish individual stars, while at the same time low mass stars are generally harder to count, due to the inability to easily detect smaller, fainter stars \citep{2009A&A...495..147A,2008ApJ...677.1278M}.  Finally, resolution becomes problematic at distances greater than $\sim$ 50 kpc, even with the  \emph{Hubble Space Telescope} (HST). For measurements of the low end of the IMF, lower--density fields can alternately be used, as the lifetimes of stars less massive than the Sun are longer than the Hubble time \citep{2000cucg.confE...7Z,2013ApJ...771...29G}. However, this method cannot be applied to the high end of the IMF, where rapid evolution depletes the Main Sequence on short timescales, depending on the stellar mass range considered.

 \begin{figure}[h!] 
   \centering
   \includegraphics[width=3.6in]{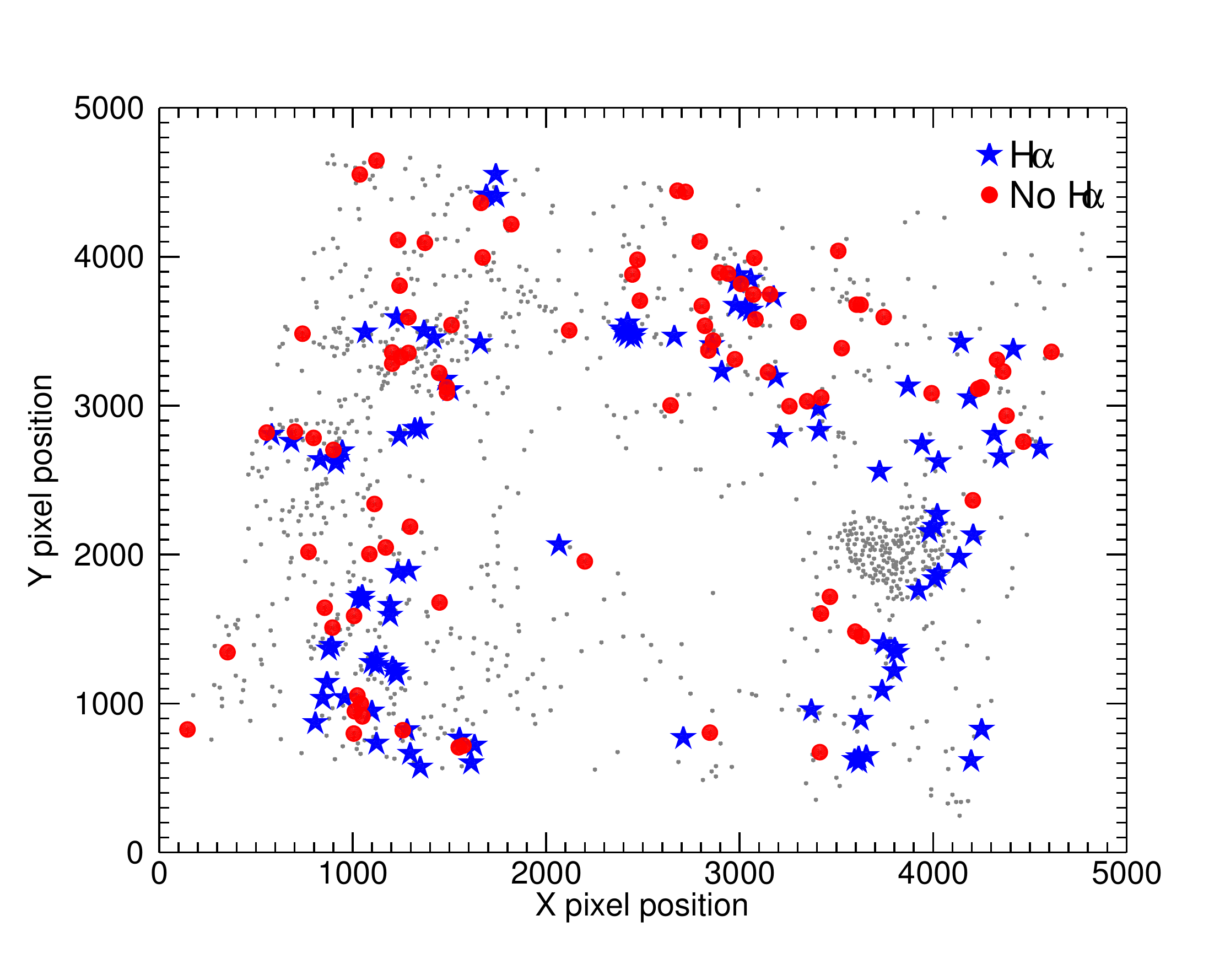}
     \includegraphics[width=3.6in]{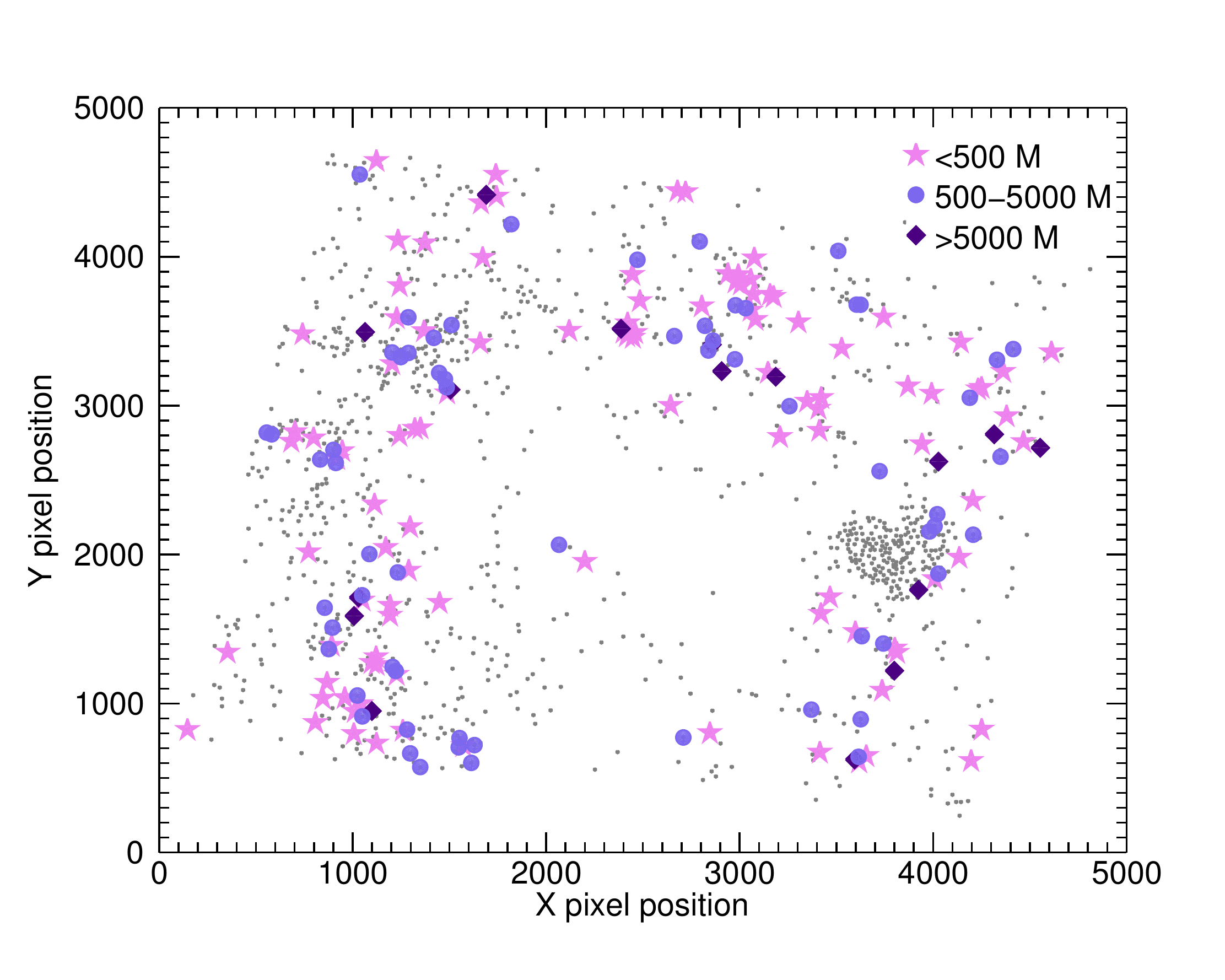}
       \includegraphics[width=3.6in]{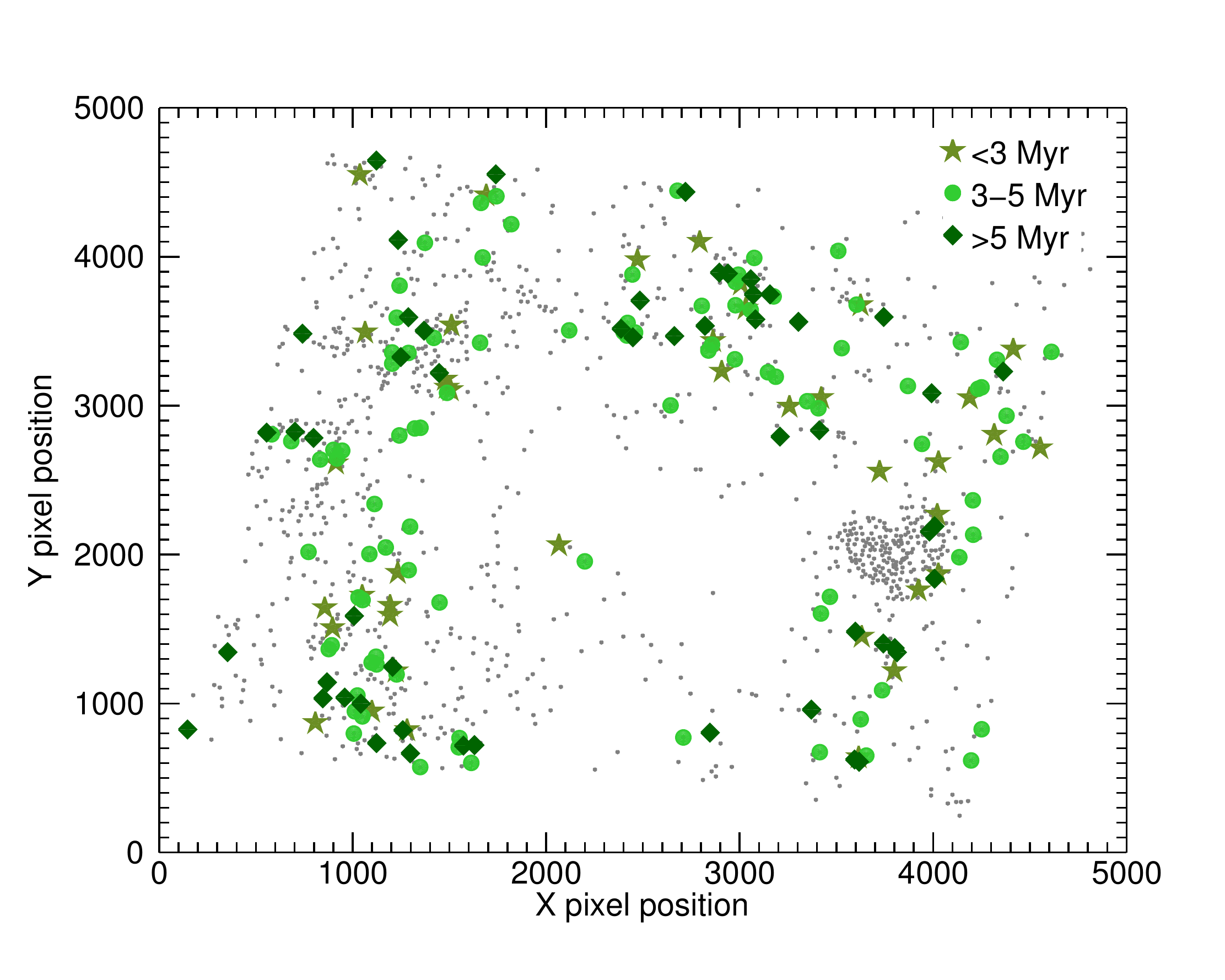}
   \caption{Positions of clusters in M83 used in this paper, subdivided by the presence of ionizing photons (top), cluster mass (middle), and age (bottom). The gray dots in each figure are from the full sample of clusters from \citet{2010ApJ...719..966C}.  }
   \label{fig:xyplots}
\end{figure}

The use of individual star counts is not necessary to constrain the upper end of the IMF, as we have demonstrated  in \citet{2010ApJ...719L.158C} and \citet{2013ApJ...767...51A}. Instead, we measure Q(H$^{0}$), the hydrogen ionizing photon rate, from the young, coeval stellar clusters which is equivalent to measuring the number of massive stars. This is an extension of the method described in \citet{2009A&A...495..479C} and relies on normalizing the ionizing photon rate to the age-independent cluster mass. \citet{2010ApJ...719L.158C}  found that there was no obvious dependence of the upper mass end of the IMF on the mass of the star cluster down to $\sim$ 10$^{3}$ M$_{\sun}$, in a pilot study done on M51a.
This result was basically confirmed by \citet{2013ApJ...767...51A}, who analyzed the nearby galaxy NGC4214, which has a star formation rate about 30 times lower than M51a. \citet{2013ApJ...767...51A} also introduced a new approach to cluster mass and age determination: the use of SLUG models \citep{2012ApJ...745..145D}, where the spectral energy distributions of single age stellar populations are produced via stochastic sampling of the stellar IMF. This treatment is a better representation of the sampling of the IMF in low-mass clusters. As shown in \citet{2012ApJ...750...60F}, `deterministic' stellar population models, in which all stars in the IMF are represented, fail to properly account for the increasing scatter in luminosity and colors of clusters below masses of $\approx$5,000~M$_{\sun}$.

In this paper we present the results of another galaxy using the method introduced by \citet{2013ApJ...767...51A}, and explore whether the results obtained in NGC4214 and M51a can be extended to other galaxies as well. For this study, we have selected a portion of the face--on, grand design spiral galaxy M83, located only 4.5~Mpc away  \citep{2003ApJ...590..256T} and shown in Figure 1. This galaxy represents an important complement to M51a, as the two galaxies are both in interaction with lower--mass companions, and have comparable SFRs; M83 has an H$\alpha$ and UV SFR of 3.3 M$_{\sun}$ yr$^{-1}$ and  3.6 M$_{\sun}$ yr$^{-1}$  respectively \citep{2005ApJ...619L..83B}. The advantage in using M83 is that this galaxy is only roughly half the distance of M51a, enabling us to push our study to clusters as light as $\sim$500~M$_{\sun}$. Compared to NGC4214, M83 offers the advantage of larger cluster numbers, implying more robust statistics. 

M83 is classified as a starburst galaxy, and has vigorous star formation in the center,
and throughout its spiral arms. This, combined with its proximity, make it an excellent
candidate for extending the study of  \citet{2013ApJ...767...51A} to a spiral galaxy. In Section 2 of this paper we will discuss the observations and cluster selection criteria, in Section 3 we will present the models and age and mass determinations, and in Section 4 we discuss the results.

\section{Data Reduction and Photometry}
HST WFC3/UVIS and WFC3/IR observations were taken as part of GO 11360 (PI: O'Connell). The observations on which we concentrate here are only of the inner region and include F225W (1800s), F336W (1890s), F438W (1880s), F487N (2700s), F555W (1203s), F657N (1484s), and F814W (1213s), shown in Figure \ref{fig:clusterpos}. Throughout the text we will refer to these as \textit{NUV}, \textit{U}, \textit{B}, \textit{H$\beta$}, \textit{V}, \textit{H$\alpha$}, and \textit{I} respectively. Each flat-fielded image was co-added, cosmic rays were removed, and corrections for distortion were made using the task MULTIDRIZZLE into a final pixel scale of 0$\arcsec$.0396 pixel$^{-1}$. At the distance of M83, the pixel scale is 0.876 pc pixel$^{-1}$. See \cite{2010ApJ...719..966C} for a full explanation of the reduction procedure.  

 \begin{figure*}[ht] 
   \centering
   \includegraphics[width=2.5in, angle=270]{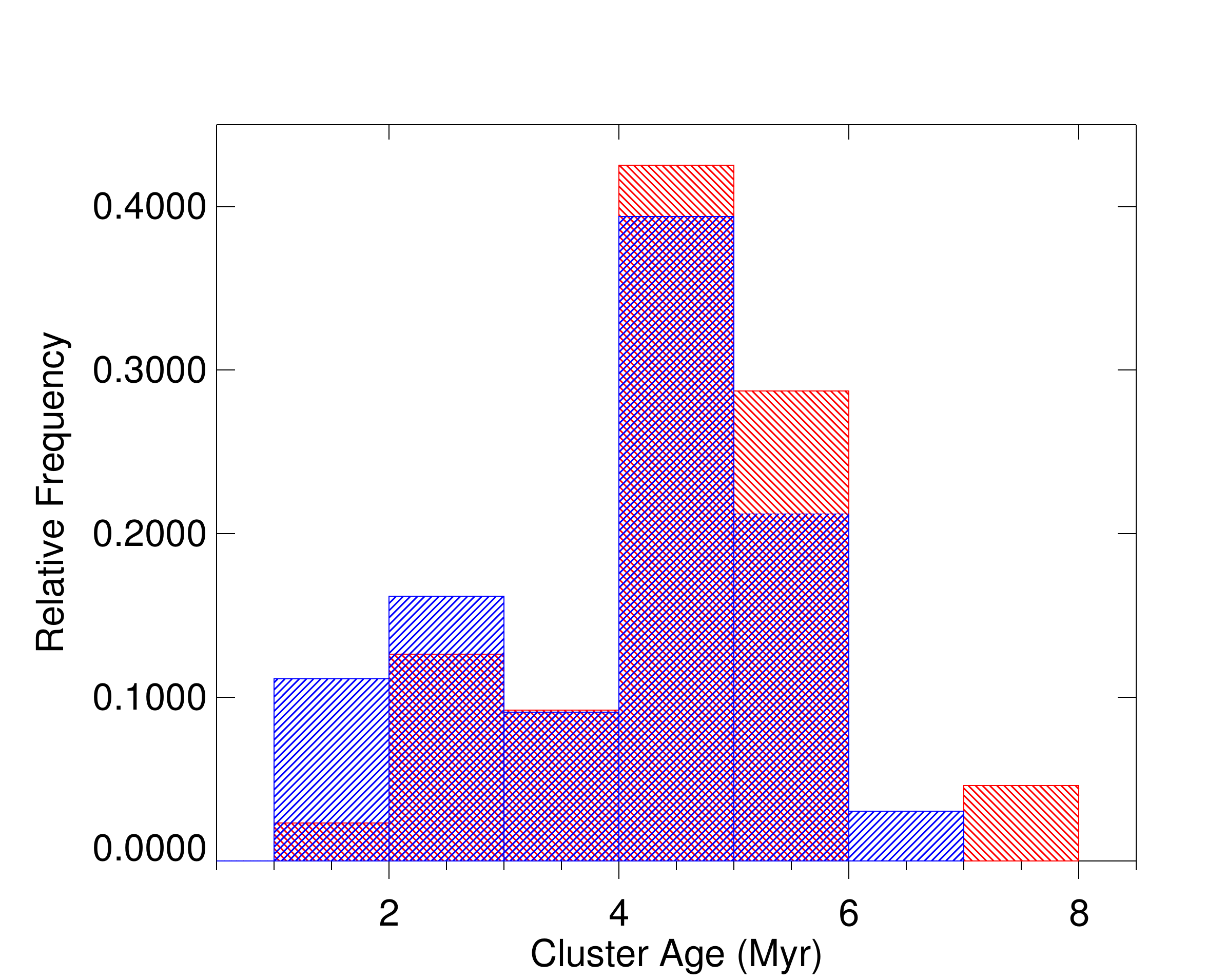} 
   \
   \includegraphics[width=2.5in, angle=270]{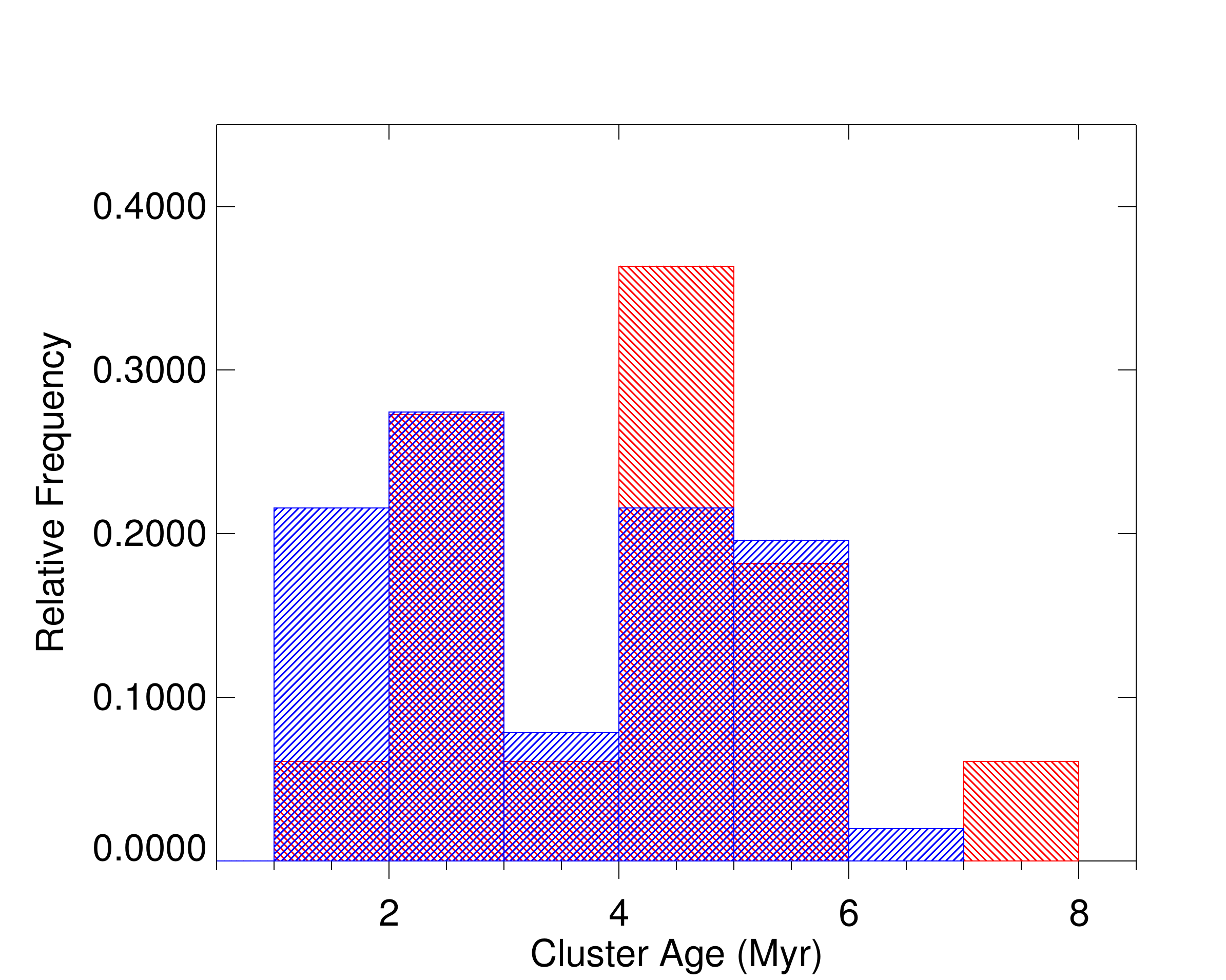}
   \caption{Histograms of relative frequencies of best fit cluster ages for all young clusters (left) and clusters with masses $>$ 500 M$_{\sun}$ (right).  The blue histogram indicates those clusters with measured H$\alpha$ emission, while the red indicates those with only upper limits. These plots exclude those clusters with masses $>$ 10$^{4}$ M$_{\sun}$ as they may not be fully explored by the SLUG models and are far more likely to have expelled their surrounding hydrogen gas at a much younger age.}
   \label{fig:agewithha}
\end{figure*}

For this paper we use the 1247 member cluster catalog from \cite{2010ApJ...719..966C}.  These are the gray dots shown in Figure \ref{fig:xyplots}.  This catalog was created using the IRAF task DAOFIND from a ``white-light" image of co-added U, B, V and I images. Aperture photometry was performed on the wide band images using the IRAF task PHOT with an aperture of 3 pixels, and a background annulus between 10 to 13 pixels.  The aperture corrections were addressed by methods relying on the concentration index ($C$, the difference in magnitudes between 3 pixel and 0.5 pixel radius).  Photometric conversion from counts to erg cm$^{-2}$ s$^{-1}$ were accomplished using the filter dependent PHOTFLAM values from the image headers.  Galactic foreground extinction of $E(B-V)$ = 0.058 \citep{2011ApJ...737..103S} was corrected using the Milky Way extinction curve from \citet{1999PASP..111...63F}.

 Due to the more extended nature of HII regions surrounding the stellar clusters, aperture sizes that scaled with the cluster mass according to the expected Str\"{o}mgren radius were used to measure the hydrogen recombination lines on the continuum-subtracted $H\beta$ and $H\alpha$+[N II] images.  The continuum-subtracted images were created by interpolating between F438W and F555W (for $H\beta$) and F555W and F814W (for $H\alpha$) and then subtracting from the F487N and F656N images respectively.  The near solar metallicty of M83 results in very little [O III]  contamination in the F555W filter, eliminating the need for iterative image subtraction.  As was done in \citet{2013ApJ...767...51A} and \citet{2010ApJ...719L.158C}, a radius of about 0.5 R$_{Stromgren}$ was selected due to the crowding of the clusters, this corresponds to between 5-30 pixels depending on the size of the cluster.  This radius is sufficiently larger than the PSFs for both the H$\alpha$ and H$\beta$ images, so there are no concerns of PSF variations. The local background was subtracted using a 3 pixel wide annulus centered on the cluster outside of the aperture radius in order to avoid contamination from other diffuse emission.  Aperture corrections were calculated from a few, very isolated sources, and were applied to the other regions. Contamination from [N II] was removed using the average galactic [N II]/H$\alpha$ ratio of 0.53 from  \cite{2008ApJS..178..247K}. The corrections for the ionized gas extinctions were measured region by region using the corresponding H$\beta$ image using the formulation in \citet{2000ApJ...533..682C} and were applied to the H$\alpha$ luminosities. 

\section{Cluster Selection}

As part of our selection criteria, we limit the acceptable cluster age range to those $<$ 8 Myr in order to keep objects in which the HII region is still density bound and the massive stars are still retained.  Over time the compact HII regions that surround the clusters expand and disperse into the ISM \citep{2011ApJ...729...78W}, and by 8 Myr the H$\alpha$ luminosity can be less than 1$\%$ of L$_{H\alpha}$ at 1 Myr \citep{1999ApJS..123....3L}.  After this time massive stars capable of producing ionizing photons ($>$ 15-20 M$_{\sun}$) also begin to disappear. An 8 Myr age limit reduces the ionizing photon rate uncertainties while at the same time ensures we retain stars which populate the upper end of the IMF. With this parameter in place, using the age-dating procedure discussed below and in \citet{2013ApJ...767...51A}, the full sample of clusters was reduced to 1/3 of the size to $\sim$ 430 members.

 There have been previous studies on the cluster population of M83 using this data  \citep{2010ApJ...719..966C, 2012ApJ...750...60F, 2012MNRAS.419.2606B}, but these papers concentrate only on larger ( $>$ 5000 M$_{\sun}$) or older ( $>$ 10 Myr) clusters. These studies also use UBVI and H$\alpha$ observations to age-date the clusters.  We do not use H$\alpha$ emission in our SED fits so as not to bias the data.  Although H$\alpha$  is important for age-dating, as indicated by \citet{2012ApJ...750...60F}, in the age and mass range we are interested in for this paper, incorporating the NUV will allow the same accuracy of fitting as the H$\alpha$ emission \citep{2013ApJ...778..138A} while avoiding bias in our sample.

Often multiple objects can ionize the same HII region, creating a situation in which it is impossible to assign a correct measurement of  L$_{H\alpha}$ to an individual cluster. Therefore in instances where clusters may share in ionizing the hydrogen gas,  we remove the objects from our sample.  This is a common occurrence in very crowded regions, including the most crowded part of the nucleus of M83. Single, large bright stars can also be problematic, and we cannot rule out the possibility that up to 30$\%$ of the objects in our lowest mass bin are in fact single stars or a tightly bound cluster of a single O star surrounded by much smaller stars.   Of the 48 members in our lowest mass bin, there are 14 objects which are likely massive stars with solar or sub-solar companions, but we can not definitively rule out the possibility of them being a single massive star except for the presence of some excess emission in the V and I bands. 

This has left us with a total of 187 clusters between the ages of 1-8 Myr, 84 of which have a best fit mass that is $\geq$ 500 M$_{\sun}$, and either have a measured H$\alpha$ luminosity (49) or have an H$\alpha$ luminosity that is non-detectable down to the 3$\sigma$ limit of 6.6 $\times$ 10$^{35}$ erg s$^{-1}$. The positions of these 187 clusters throughout the galaxy based on the presence of H$\alpha$, masses, and ages are all shown in Figure \ref{fig:xyplots}. These three panels also illustrate that our clusters populate the spiral arm quite thoroughly. 

\section{Models and Analysis}
\begin{figure*}
   \centering
   \includegraphics[width=2.5in,angle=270]{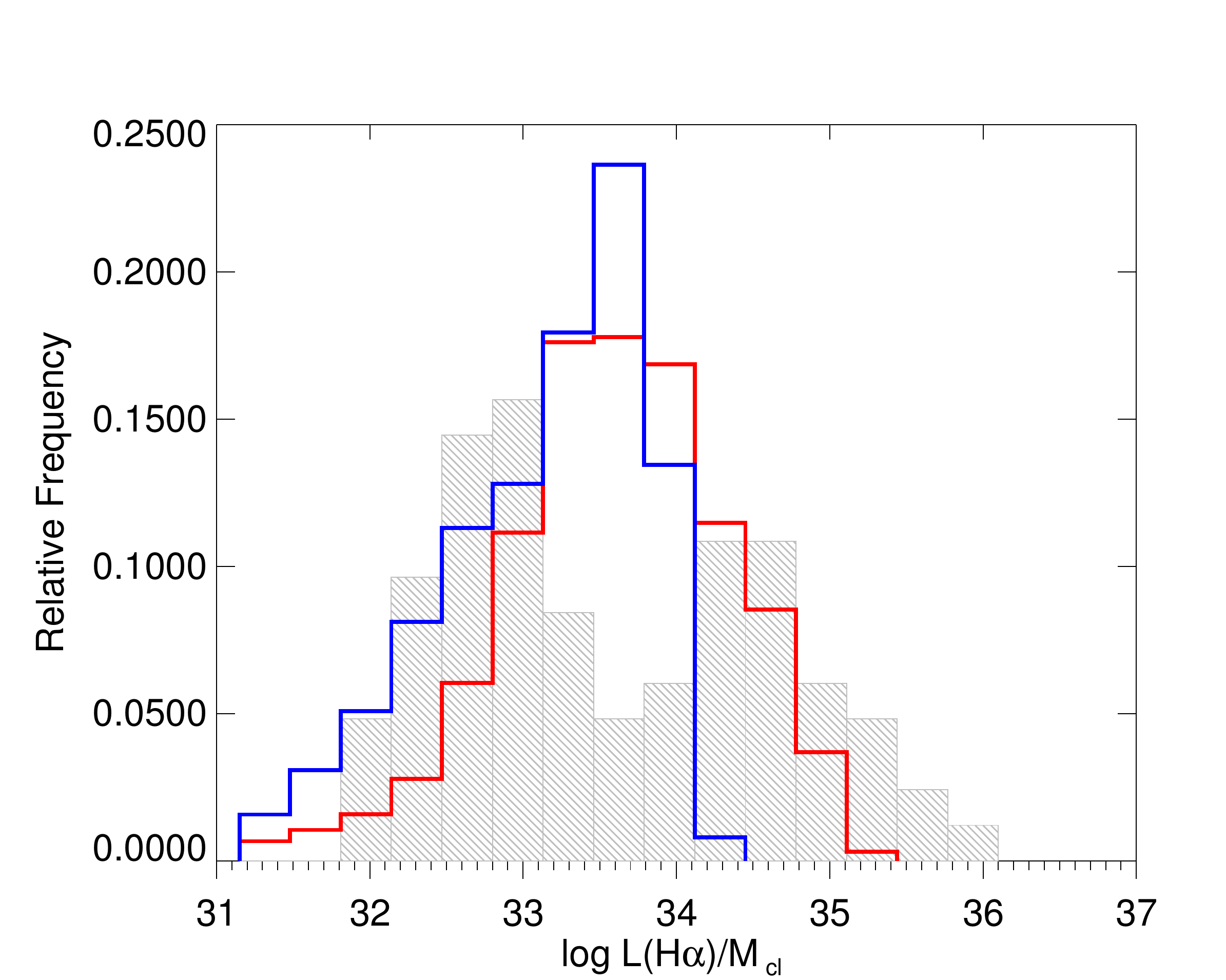} 
    \includegraphics[width=2.5in, angle=270]{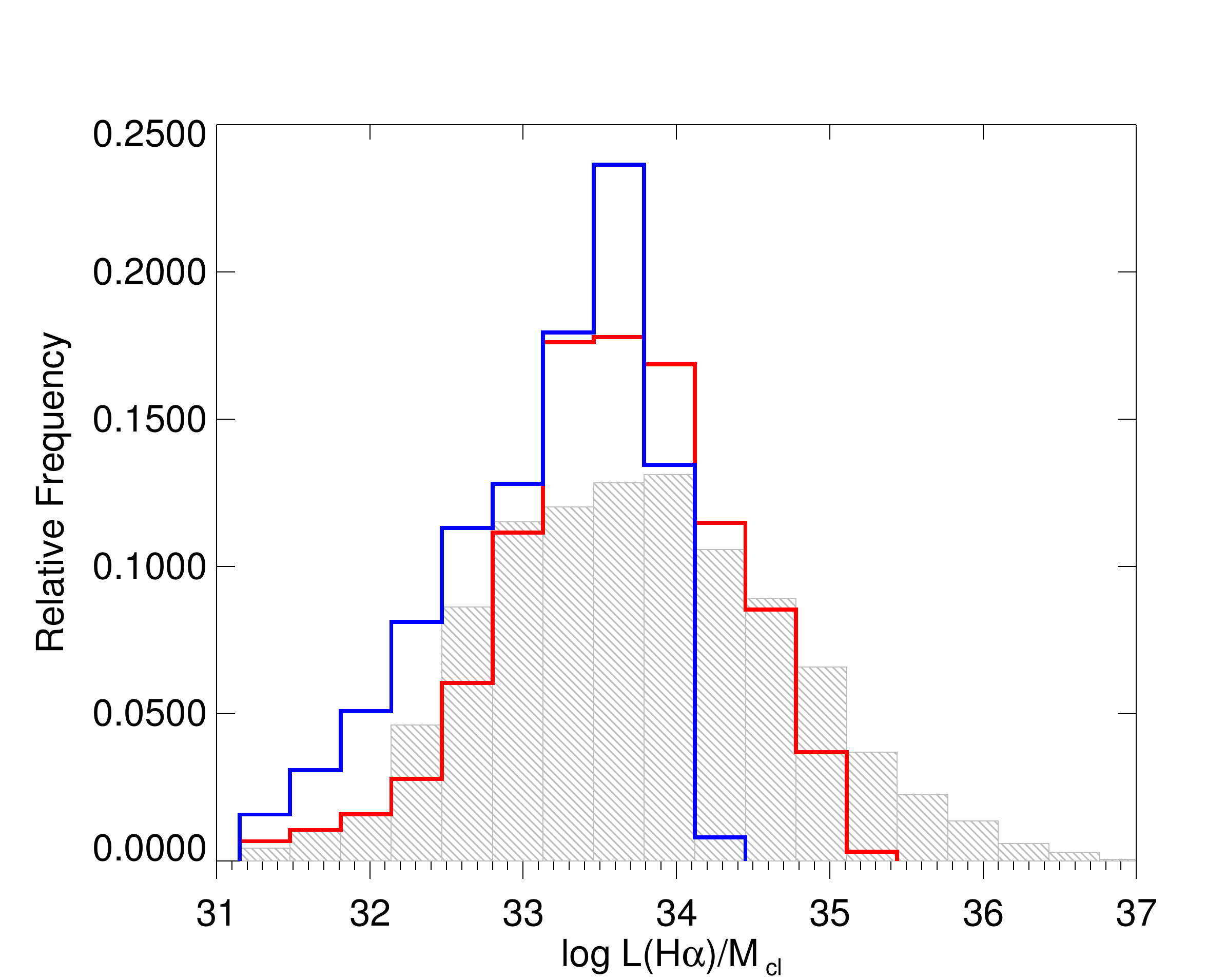}
   \includegraphics[width=2.5in,angle=270]{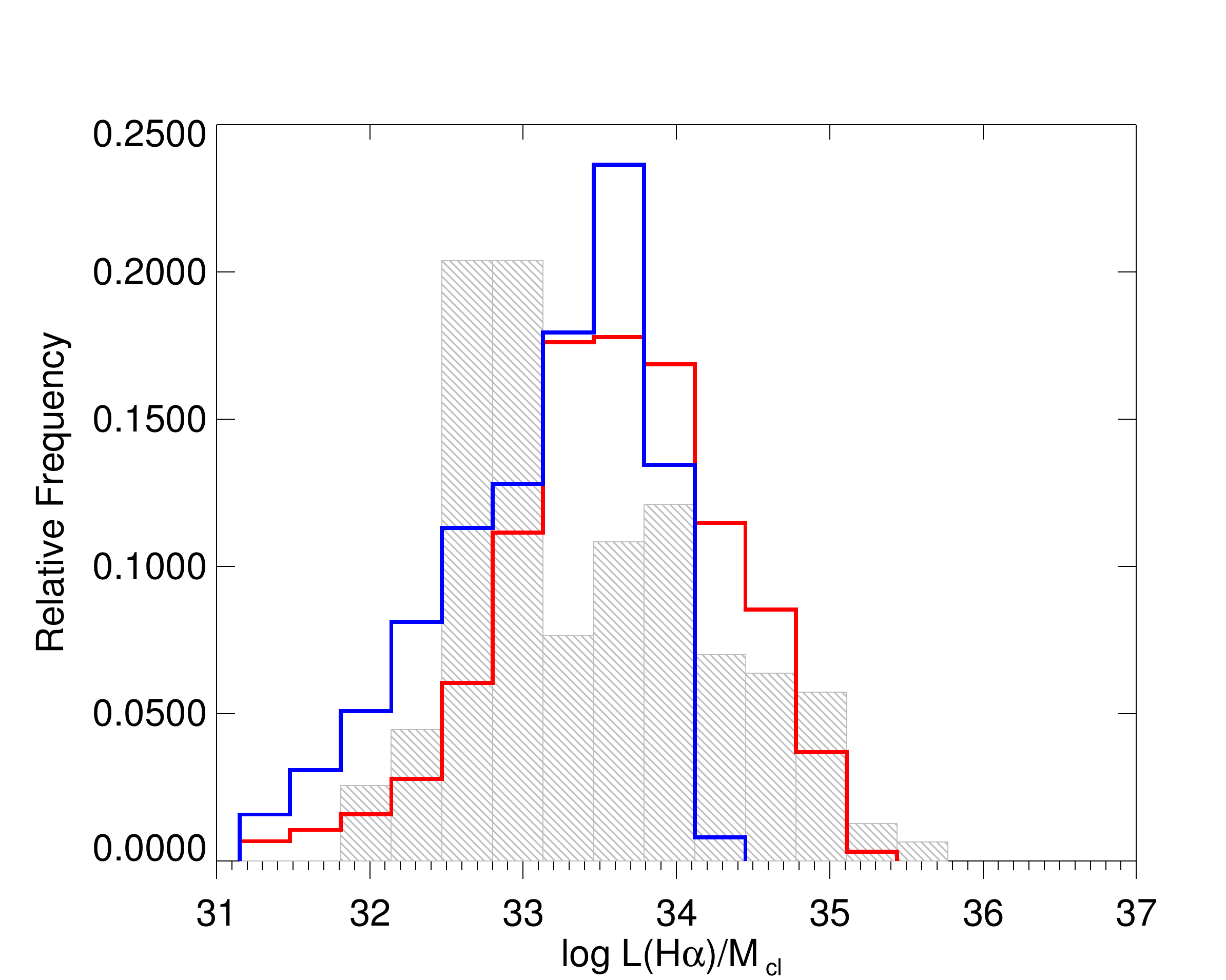}
    \includegraphics[width=2.5in,angle=270]{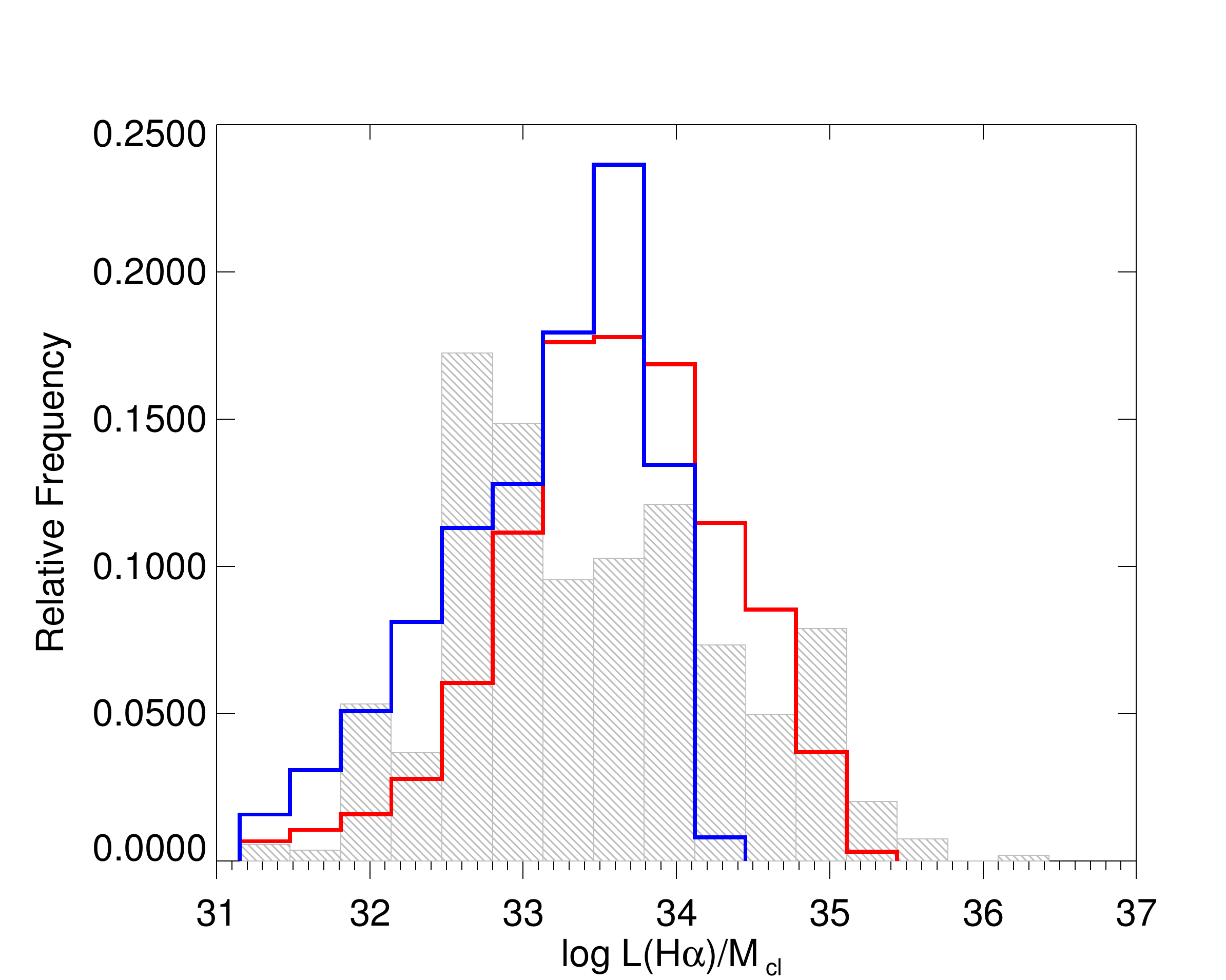}
   \caption{Histograms of L$_{H\alpha}$ /M$_{cl}$ from SLUG models for a fully sampled IMF (red) and a model truncated at 30 M$_{\sun}$ (blue) for L$_{H\alpha}$ /M$_{cl}$ plotted against clusters from M83 whose masses were determined from SLUG models with a maximum stellar mass of 120 M$_{\sun}$ (top) and SB99 models with a maximum stellar mass of 30 M$_{\sun}$ (bottom). In all cases, clusters with only measured H$\alpha$ upper limits are given the 3$\sigma$ limit of 6.6 $\times$ 10$^{35}$ erg s$^{-1}$ as the luminosity value. The left panels only include the single best-fit mass.  The right panels take each solution with a $\chi ^{2}$ $<$ 1 and gives it equal weighting, which in practice creates 53629 and 545 distinct entries for SLUG (top) and SB99 (bottom) respectively.}
   \label{fig:histo}
\end{figure*}

In an extension of \citet{2013ApJ...767...51A}, we have computed the L$_{H\alpha}$/M$_{cl}$ of young ($<$ 8 Myr), stellar clusters in Field 1 of M83 down to $\sim$ 500 M$_{\sun}$ using stochastic models and both a canonical and truncated IMF.  To determine the ages and masses of the clusters we have used the photometry of the 5 broad-bands discussed in Section 2, but choose not to include the  H$\alpha$ narrow-band filter in the age-dating as the presence or absence of H$\alpha$ emission may bias our sample. The best fit model-derived masses of clusters with ages less than 8 Myr are then binned into three distinct mass bins of roughly equal masses.  Within these mass bins the masses are summed and the H$\alpha$ luminosities are summed to determine the  L$_{H\alpha}$/M$_{cl}$ ratios for various cluster masses ($<\frac{L_{H\alpha}}{M_{cl}}>  =\frac{\sum_{i}L_{H\alpha i}}{\sum_{i}M_{cli}}$). These ratios are then compared to predicted models with two different assumptions about the IMF, a universal one and one in which the upper mass limit is a function of the cluster mass, in a formulation similar to that of \citet[see Discussion]{2003ApJ...598.1076K}.

As with NGC 4214, we used both the Starburst99  \citep[hereafter SB99]{1999ApJS..123....3L,2010ApJS..189..309L, 2014ApJS..212...14L} deterministic models, and the SLUG \citep[Stochastically Lighting Up Galaxies]{2012ApJ...745..145D} stochastic models. Stochastic models are important, especially at the low cluster mass end since as cluster size decreases, the influence of the massive star or stars in the cluster becomes much more prominent.  Both sets of models use a Kroupa IMF between 0.08-120 M$_{\sun}$ \citep{2001MNRAS.322..231K}, and assume that the clusters form in a single instantaneous burst. Although the metallicity of M83 is roughly 1.5$\times$ solar \citep{2002ApJ...572..838B}, the combination of metallicities available in SB99 and the super-solar to solar metallicity gradient in the galaxy  has made the Padova AGB tracks with z=0.02 the best choice for both models. Additionally, both \citet{2011A&A...532A.147L} and \citet{2012ApJ...753...26K} have shown that the use of solar or 1.5$\times$ solar metallicty in M83 does not produce significant differences in the age determination.

 For the truncated SLUG models, where the maximum mass was only allowed to be 30 M$_{\sun}$, we had access to only models that used a Salpeter IMF; we rescaled these to match the Kroupa IMF.  We do not expect this difference to impact our results, as the two IMFs only differ below  0.5 M$_{\odot}$, i.e., well below the stellar mass range of interest in this study. The SLUG models include about 5000 cluster templates with ages from 1 Myr to 20 Myr with a cluster mass of 1 $\times$ 10$^{3}$ M$_{\sun}$.  As noted in \citet{2013ApJ...767...51A}, the uncertainties introduced using 1 $\times$ 10$^{3}$ M$_{\sun}$ models for less massive and more massive clusters is small, and is already encompassed by the uncertainties generated within the data and the 1 $\times$ 10$^{3}$ M$_{\sun}$ models themselves. Additionally, while the number of SLUG models that can be generated greatly exceeds the 5000 used here, we have found that we are getting consistent fits with low $\chi ^{2}$ values with the reduced amount of models. It should also be noted that SLUG models do not allow for binarity, but a recent study by \citet{2012MNRAS.422..794E} indicates that at masses $\geq$ 10$^{3}$ M$_{\sun}$, the scatter in L$_{H\alpha}$/M$_{cl}$ is the same between models that use single stars and those that introduce binaries, so this should not introduce additional uncertainties.

We have utilized a reduced $\chi ^{2}$ fitting technique between both the SLUG and SB99 models and the cluster photometry to determine the age, mass, and extinction of each cluster.  For each cluster an SED of the five photometric data points were compared to both SLUG and SB99 models spanning the complete reddening range between 0$\leq$$E(B-V)$$\leq$1.00 in intervals of 0.05.  It is important here to point out that there is no single solution for the age and extinction of the cluster, but instead there is a range of best fits which could produce the model fit. By using the large range of ages and extinction consistent with the model fits, the actual mass distribution has been extended over a range of values and only produces a peak at the most probable value. Therefore, we allowed all  $\chi ^{2}$  values less than one, as was done in \citet{2003MNRAS.345..161P} and with NGC 4214, and include all ages, extinctions, and therefore corresponding masses within that range. This means that for each cluster there is more than one acceptable fit, but only one best fit (the one with the lowest $\chi ^{2}$ value). As an example, if we take one of our lowest mass clusters, the best fit is a mass of 574 M$_{\sun}$ with an age of 4.25 Myr and an E(B -V) = 0.05.  The average value of these parameters from all of the fits with a reduced  $\chi ^{2}$ less than 1 are 550 M$_{\sun}$ with an age of 4.2 Myr and an E(B-V) = 0.045, extremely similar to the best fit mass.  Plots of this cluster, like those in Figure 3, bottom in \citet{2013ApJ...767...51A}, show that all the models agree that E(B-V) must be less than 0.05, and that there is a global minimum between 4-5 Myr, and 300-700  M$_{\sun}$, where all $\chi ^{2}$ values range from 0.18-0.4.  There are  two local minima at 1.5-3 Myr/200-600 M$_{\sun}$ and 8 Myr/700-1500 M$_{\sun}$ but these all have $\chi ^{2}$ values $>$ 0.4, implying that these solutions are quite unlikely.  This sort of behavior is seen in all of the other cluster fits as well.When all acceptable fits are used, particularly for uncertainty purposes, each fit is given equal weighting.  Therefore instead of 84 entries for the best fit, there are over 50,000 entries which satisfy the condition of $\chi ^{2}$ $<$ 1. This is discussed more in Section 5. 

\section{Discussion}

\begin{figure*}[] 
   \centering
   \includegraphics[width=6.3in]{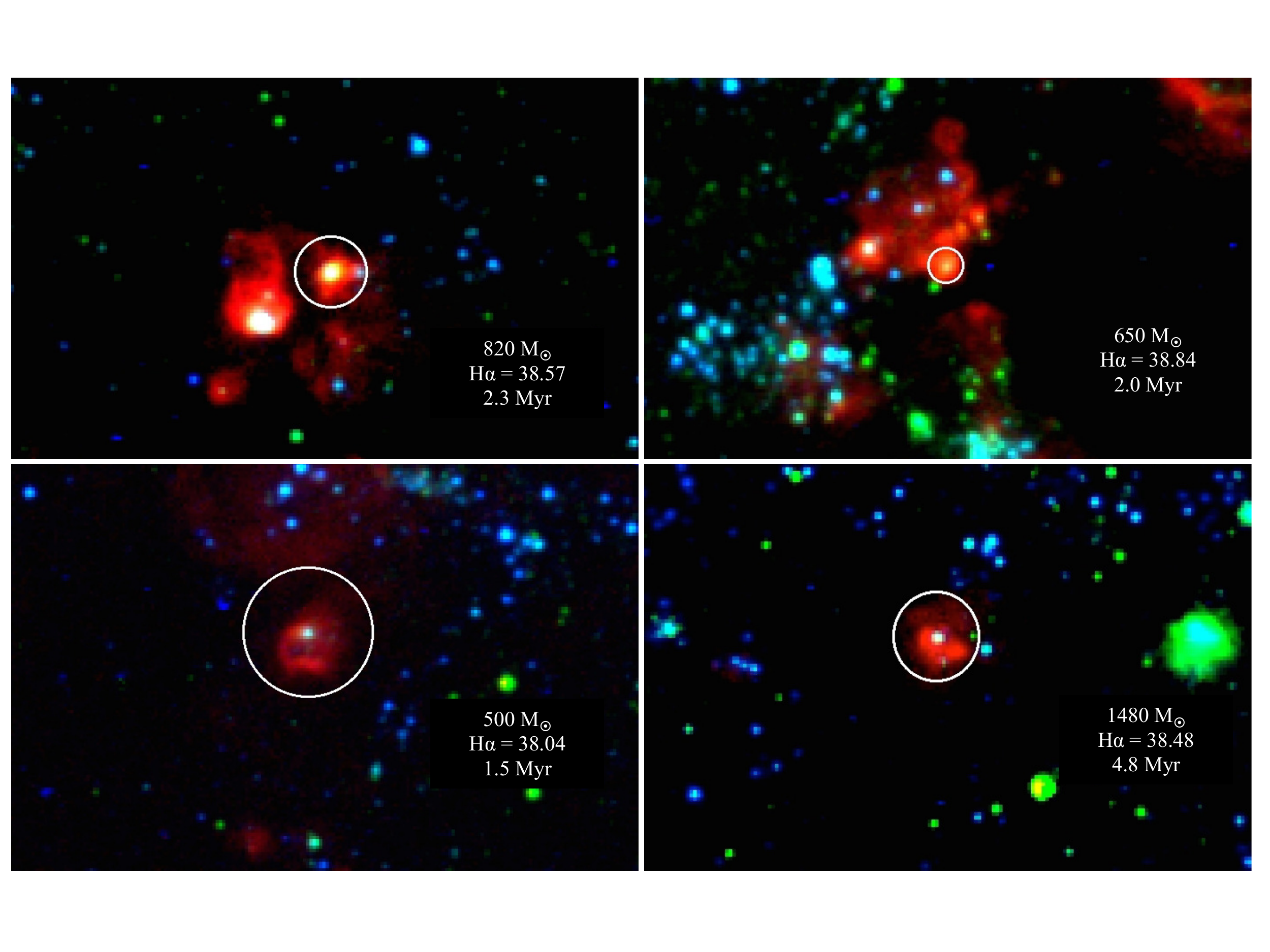} 
   \caption{F225W (blue), F555W (green), and continuum subtracted H$\alpha$ (red) composite images of a sample of low mass clusters with high ionizing photon rates. The mass, log(L$_{H\alpha}$), and age of each cluster is indicated on each image, as well as the size aperture used in the H$\alpha$ photometry (white circles).   The RA and DEC for each cluster are (clockwise from top left): 13$^{h}$37$^{m}$09$^{s}$.02, -29$^{\circ}$52$^{\prime}$09$^{\prime\prime}$.31; 13$^{h}$37$^{m}$07$^{s}$.70, -29$^{\circ}$51$^{\prime}$11$^{\prime\prime}$.26;  13$^{h}$37$^{m}$09$^{s}$.56,  -29$^{\circ}$52$^{\prime}$23$^{\prime\prime}$.76; and 13$^{h}$37$^{m}$09$^{s}$.70,  -29$^{\circ}$52$^{\prime}$43$^{\prime\prime}$.47. }
   \label{fig:clustermontage}
\end{figure*}

\subsection{ H$\alpha$ emission with age}

Histograms of the best fit ages for all clusters younger than 8 Myr with masses $<$ 10$^{4}$ M$_{\sun}$ (left), and all clusters with masses between 500 - 10000 M$_{\sun}$ (right) are shown in Figure \ref{fig:agewithha}.  In both plots those clusters with detected H$\alpha$ are shown in blue and those with only upper-limit detections in H$\alpha$ are shown in red. Unlike the NGC 4214 results which clearly indicated all clusters with ages that lie between 6-8 Myr have non-detections in H$\alpha$ and all clusters younger than 4.5 Myr are detected in H$\alpha$, there only seems to be a lack of H$\alpha$ detections above 7 Myr.  The differences could possibly be caused by the further distance or higher metallicity of M83; although the data is still consistent with very little H$\alpha$ detection in clusters older than 6 Myr.

According to \citet{2012MNRAS.423.2933R}, the leakage of ionizing photons is expected for those HII regions with ages greater than 4 Myr, which is where we see the greatest number of clusters without H$\alpha$ emission in our M83 sample. Some of the clusters with H$\alpha$ emission do have ages greater than 5 Myr, but we do need to be aware of the fact that especially in clusters that may contain only one or two extremely massive stars they may not live long enough to produce ionizing photons out to 8 Myr.  For example, the lifetime of a 35 M$_{\sun}$ star is roughly 5 Myr, while a 15 M$_{\sun}$ star may live 15 Myr and a 120 M$_{\sun}$ star only 2 Myr.  If there is only one massive star in these smaller clusters, the L$_{H\alpha}$  may be more sensitive to the age.  As a lower limit, if we use a fully populated IMF to estimate the number of 15, 35, and 120 M$_{\sun}$ stars expected in our cluster mass range of 500 - 10$^{4}$ M$_{\sun}$, the result is between 3 - 60, 0-5, and none respectively.  Stochasticity is therefore extremely important, especially in the lower mass clusters where it is unlikely to find stars massive enough to power an HII region.

\subsection{ IMF Variations}

In the cluster--mass--dependent upper mass limit formulation of \citet{2010MNRAS.401..275W}, a M(max)$_{*}$ -- M$_{cl}$ relation is proposed in which the most massive star in a cluster is limited by the mass of the parent cluster.  For instance in a 1000 M$_{\sun}$ cluster no stars more massive than 35 M$_{\sun}$ would be present, and an ensemble population of 10$^{3}$ M$_{\sun}$ clusters would never fully populate the IMF.  This truncated formulation is represented as the blue histogram shown in the panels of Figure \ref{fig:histo}. Conversely, in a purely stochastically populated IMF, 100 10$^{3}$ M$_{\sun}$ clusters would contain the same numbers and masses of stars as one 10$^{5}$ M$_{\sun}$ cluster, and that both would represent a fully sampled IMF \citep{2001ASPC..243..255E,2006ApJ...648..572E}.  In recent years, the M(max)$_{*}$ -- M$_{cl}$ relation has been allowed by the same authors to include some stochasticity \citep{2014MNRAS.441.3348W}, although it is still the case that only for a universal IMF a star cluster can be over--luminous in H$\alpha$ relative to what would be expected for its mass.  This can happen if a low-mass cluster is, for purely random reasons, particularly rich in massive stars.  Conversely, in a formulation in which there is a cluster--mass--dependent truncation to the IMF, clusters will be unlikely to be H$\alpha$--over--luminous.  

We have found in our study of M83 that indeed there are low-mass clusters with large ionizing photon rates.  For example, the clusters shown in Figure  \ref{fig:clustermontage} have masses ranging from 500 - 1500 M$_{\sun}$ with corresponding  H$\alpha$ luminosities between  1.1 - 6.9 $\times$ 10$^{38}$ erg s$^{-1}$.  These, and similar objects, are responsible for the tail rightward of  L$_{H\alpha}$/M$_{cl}$ $=$ 34 in the histograms of L$_{H\alpha}$/M$_{cl}$ shown in Figure \ref{fig:histo}.  When masses are calculated using the fully populated SLUG models (top) there is an under-abundance of low L$_{H\alpha}$/M$_{cl}$ values and an over-abundance of high values, in comparison to the predictions of a truncated model (blue).  While the agreement between the fully populated IMF model (red) and our data may not be perfect, the observations still indicate that there is no obvious decrease in L$_{H\alpha}$/M$_{cl}$ values for decreasing M$_{cl}$ (Figure \ref{fig:histo}). The high L$_{H\alpha}$/M$_{cl}$ tail in low M$_{cl}$ clusters exists even when we only allow our cluster masses to be measured using models with a maximum stellar mass of 30 M$_{\sun}$ (bottom). This tail in M83 is not very different from that reported for NGC4214 in \citet{2013ApJ...767...51A}. This result is striking in that it disagrees with a simple M(max)$_{*}$ -- M$_{cl}$ relation reported for young star clusters in the Milky Way \citep{2010MNRAS.401..275W, 2013MNRAS.436.3309W}.  We should note that the star formation rates in NGC4214 and M83 bracket that of the Milky Way, thus the disagreement is real. Ways to reconcile the different results may require investigating the consequences of using different methods to measure the cluster masses, and a careful analysis of what uncertainties each method carries; for a discussion of the problems with measuring cluster masses in the Milky Way see \citet{2014arXiv1402.0867K}.

\begin{figure}[h] 
   \centering
   \includegraphics[width=3.6in]{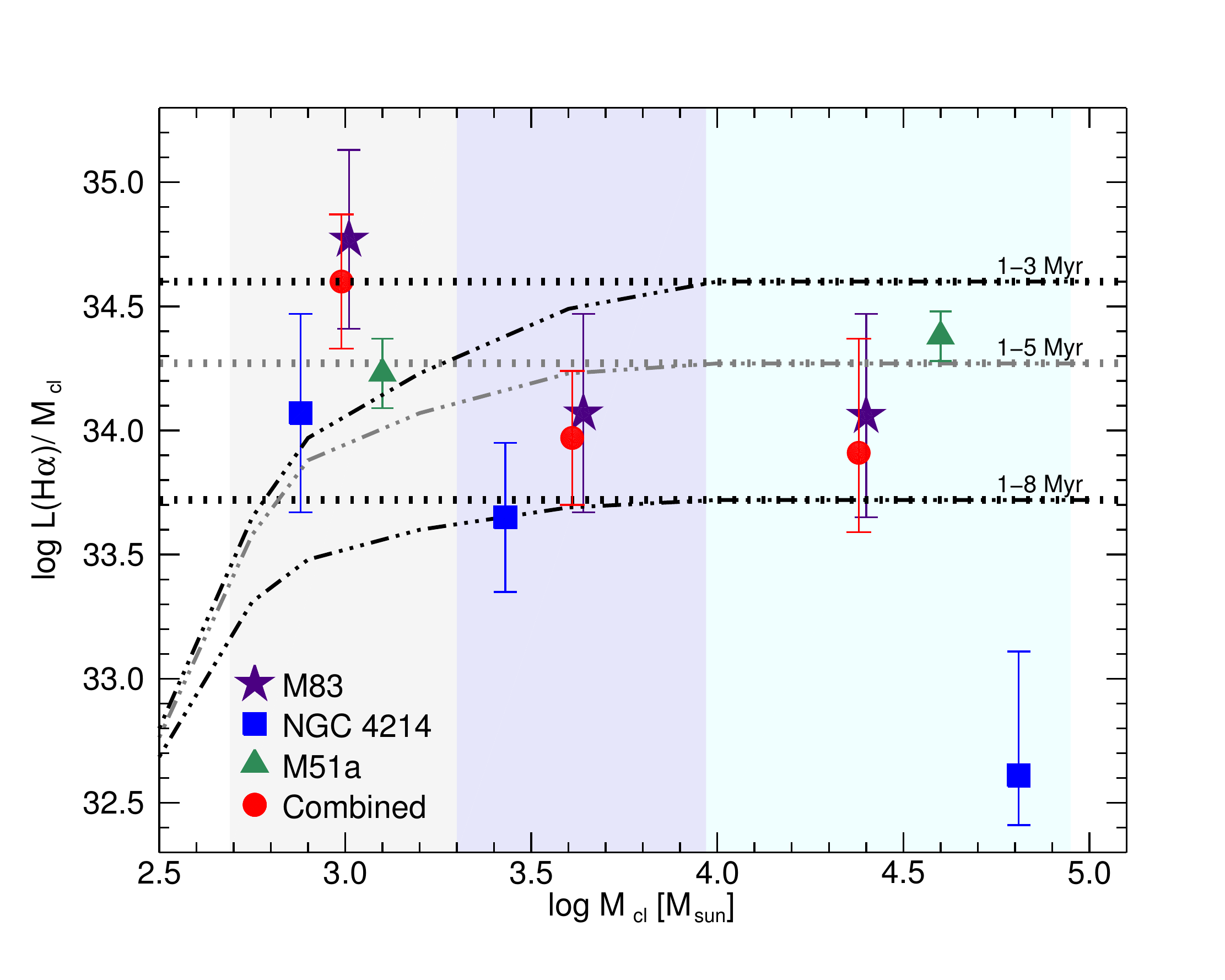} 
   \caption{ Location of L$_{H\alpha}$ /M$_{cl}$ for mass bins in M83, NGC 4214 \citep{2013ApJ...767...51A}, M51 \citep{2010ApJ...719L.158C}  and combined bins for M83 and NGC 4214 for clusters $<$ 8 Myr. All galaxies have been normalized to the metallically of M83 ( Z = 1 Z$_{\sun}$). Dotted lines are the expected L$_{H\alpha}$ /M$_{cl}$ for an universal IMF for various age ranges while the dash-dotted line is for a cluster--mass--dependent upper mass limit \citep{2003ApJ...598.1076K,2010MNRAS.401..275W}  where the most massive star in a cluster is determined by cluster mass. Each shaded region indicates the mass range for each mass bin (500-2000 M$_{cl}$, 2000-9000 M$_{cl}$, and 9000+ M$_{cl}$).}
   \label{fig:binnedplot}
\end{figure}

In a M(max)$_{*}$ -- M$_{cl}$ relation, the summation of the total ionizing flux from the small clusters divided by the total cluster mass should be much lower than the ionizing flux from a single large cluster divided by its mass and as cluster mass decreases there is a deviation from the ratio of ionizing photons to mass expected by a universal IMF (Figure \ref{fig:binnedplot}, dashed-dotted line). Whereas in an universal IMF scenario this summed ratio would be consistent with that of a single large cluster.  Of course an universal IMF predicts as a whole clusters $\leq$ 500 M$_{\sun}$ will mostly produce low H$\alpha$ luminosities.  In fact, \cite{2010A&A...522A..49V} estimates that only 20$\%$ of 100 M$_{\sun}$ clusters will have stars large enough to create an H II region. There will be some low mass clusters that do produce a large ionizing continuum from the odd star well over 20 M$_{\sun}$ (case in point, Figure \ref{fig:clustermontage}), so the effects can be averaged out if the sample size is large enough. We have therefore minimized both the observational uncertainties and the stochastic effects by summing the L$_{H\alpha}$ and masses of all of the small clusters into one data point.

The data have been combined into three mass bins (see the three shaded regions in Figure \ref{fig:binnedplot}), each with a mean mass of 9.8  $\times$ 10$^{2}$ M$_{\sun}$, 1.8 $\times$ 10$^{3}$ M$_{\sun}$, and 2.8 $\times$ 10$^{4}$ M$_{\sun}$.  The error bars have been calculated by adding in quadrature the individual mass and luminosity uncertainties of each cluster fit.  The expected average L$_{H\alpha}$/M$_{cl}$ from a solar metallicity SB99 model that is fully populated up to 120 M$_{\sun}$ has also been plotted in Figure \ref{fig:binnedplot}. The top dashed line is for the average model between 1-3 Myr, the gray dashed line for ages between 1-5 Myr, and the bottom dashed line shows an averaged 1-8 Myr model.  The expected range for a M$_{*}$-- M$_{cl}$ model where the most massive star in the cluster is a function of cluster mass \citep{2010MNRAS.401..275W} is shown in the dashed-dotted lines also averaged between 1-3 Myr (top), 1-5 Myr (gray) and 1-8 Myr (bottom).  The data from other galaxies have been normalized to the same metallicity of 1 Z$_{\sun}$ for accurate comparison.  Even if the L$_{H\alpha}$ estimations from various stellar evolution models are taken into account, this would serve only to move the lines uniformly up or down, and would not change the trend of the plot which shows, within uncertainties, a consistent L$_{H\alpha}$/M$_{cl}$ over all mass ranges.  It is clear both from the individual M83 measurements and combined M83 + NGC 4214 that particularly in the low mass regime, we do not find an absence of ionizing photons, and therefore massive stars must be forming in these low-mass clusters. The M51 data have not been combined with the other galaxies as it was only analyzed using SB99 models \citep{2010ApJ...719L.158C}. We must, of course, be mindful that this result, while seen in all of our galaxies, is only at the 2-3 $\sigma$ level, and a larger sample size is needed for the most conclusive results possible.

As an added result, the presence of O stars in these low mass clusters imply that they do not seem to terminate star formation.   The fact that we recover the same average number of O stars per unit cluster mass as predicted by the randomly sampled IMF suggests that if we have low mass clusters forming a single O star, then we must also have high mass clusters forming multiple O stars. For example, one O star in a 500 M$_{\sun}$ cluster would need to be balanced out by two O stars in a 1000 M$_{\sun}$.  The presence of multiple O stars suggests that the first massive star forming in the larger clusters does not impede the formation of the second O star and so on, and that O stars do not terminate star formation in clusters of any mass.  This implies two possible scenarios which allow the O stars to clear natal gas in a non-destructive way; either the cluster formation timescale is rapid enough that gas ejection by O stars happens too late to modify the final cluster mass, or that mass ejection by O stars is not sudden.  If instead, massive stars lose mass through a steady wind, star formation will be terminated at some later time, but not in a way that would be reflected in the total stellar mass formed.

\section{Conclusions}

Using the methods of \citet{2010ApJ...719L.158C} and \citet{2013ApJ...767...51A}, we have probed the presence of massive stars in a portion of the spiral galaxy M83 via the ratio of the luminosity of the ionizing photons normalized to the mass of the cluster in an effort to constrain the upper end of the IMF.  The final sample of 84 clusters with masses $>$ 500 M$_{\sun}$ and ages $<$ 8 Myr indicate that even at masses $\sim$ 10$^{3}$ M$_{\sun}$, there does not seem to be a deviation from the expected ionizing flux of an universal IMF up to 120 M$_{\sun}$.  As an extension of this, we have combined this data with the 52 clusters of NGC 4214 from \citet{2013ApJ...767...51A} corrected for metallicity differences for a more robust sample. In all instances, clusters with best fit masses down to 500 M$_{\sun}$ have a L$_{H\alpha}$/M$_{cl}$ ratio that is consistent with that predicted by an universal IMF.  This study is also supported by the results of \citet{2012ApJ...750...60F}, who use the same cluster catalog but extend their study to older cluster ages. 

Our analysis of this sample suggests that the young clusters seen in M83 can not be sufficiently explained by a truncated IMF (one in which the maximum M$_{*}$ is a function of M$_{cl}$), which would result in the maximum stellar mass in a cluster of 10$^{3}$ M$_{\sun}$ being no greater than 35 M$_{\sun}$ \citep{2010MNRAS.401..275W}.  We can not discount that it is possible if recent findings of  \citet{2014ApJ...780...27P} are invoked, where the M$_{max}$ -- M$_{cluster}$ relation can be described more thoroughly as range of maximum masses and not one single value, this relation can be appropriately applied.  In this altered M$_{max}$ -- M$_{cluster}$ model, the mass range for the most massive star in a 500 M$_{\sun}$ cluster is 15 - 72 M$_{\sun}$, a range where all stars are capable of producing ionizing photons.

We have concluded that the summation of individual young clusters in this portion of M83 is better interpreted as an universal IMF without a truncation of massive stars.  This is in agreement with the recent paper by \citet{2011ApJ...741L..26F}, who investigated the integrated properties of individual galaxies using the SLUG models.  Furthermore, we have found not only a disagreement between the observations presented here and that of the M$_{*}$ -- M$_{cl}$ relation of \citet{2010MNRAS.401..275W}, but also that the data point to a stochastically-sampled IMF with an upper mass limit consistent with a standard \citet{2001MNRAS.322..231K} IMF. Combination of this data with that of NGC 4214 strengthen this conclusion, and do not present a compelling reason for excluding an universal IMF at the high end. If anything we have found that low mass clusters appear to have exactly the opposite behavior: to be higher than expectations from both truncated and universal IMFs.

\acknowledgements{ We would like to thank the referee for contributing helpful and insightful comments. J.A. and D.C. acknowledge partial support for this study from the grant associated with program \# GO-11360 (P.I.: R.W. O'Connell), which was provided by NASA through the Space Telescope Science Institute, which is operated by the Association of Universities for Research in Astronomy, Inc., under NASA contract NAS 5-26555. MRK acknowledges support from NASA through Hubble Award Numbers AR-13256 and GO-13364, ATP grant NNX13AB84G, and TCAN grant NNX14AB52G, and from the NSF through grant AST-0955300.}

\end{document}